\documentclass[12pt]{iopart}

\usepackage{iopams}  
\usepackage{graphicx}
\usepackage{amssymb}

\newcommand{\la}{\label}
\newcommand{\be}{\begin{equation}}
\newcommand{\ee}{\end{equation}}
\newcommand{\bea}{\begin{eqnarray}}
\newcommand{\eea}{\end{eqnarray}}

\newcommand{\p}{\partial}

\def\la{\label}
\def\be{\begin{equation}}
\def\beq{\begin{equation}}
\def\eeq{\end{equation}}
\def\ee{\end{equation}}
\def\bea{\begin{eqnarray}}
\def\eea{\end{eqnarray}}
\def\p{\partial}

\begin{document}

\title{Generic critical points of normal matrix ensembles}

\author{Razvan Teodorescu}

\address{Physics Department, Columbia University, 538 West 120th Street, Mail Code 5293, New York, NY 10027}
\ead{rteodore@phys.columbia.edu}
\begin{abstract}
The evolution of the degenerate complex curve associated with the ensemble at a generic critical point is related to the finite time singularities of Laplacian Growth. It is shown that the scaling behavior at a critical point of singular geometry $x^3 \sim y^2$ is described by the first Painlev\'e transcendent. The regularization of the curve resulting from discretization is discussed.  
\end{abstract}

\submitto{\JPA}


\section{Introduction}

Random Matrix Theory, introduced to theoretical physics by Wigner and Dyson \cite{WD1,WD2} more than 60 years ago, has recently seen important new developments prompted by its application to several different problems. Of particular interest are generalizations to ensembles involving two independent
matrices, such as 2 hermitian random matrix theory (2HRM) and normal random matrix theory (NRMT). Rigorous results obtained in 2HRM over the last few years \cite{2HRM1,2HRM2,2HRM3,2HRM4,2HRM5,2HRM6,2HRM7} have led to important relations concerning  bi-orthogonal polynomials, the Riemann-Hilbert problem, and the Kadomtsev-Petviashvilii (KP) hierarchy of integrable differential equations. Results of similar nature have been obtained for normal matrix ensembles, sometimes with simple geometric interpretations \cite{US1,US2,US3,US4,US5,US6,US7}, relevant to conformal maps in two dimensions \cite{CM1,CM3,CM4}. 

Critical points of hermitian random matrix ensembles have been studied intensively because of their important relations to 2D quantum gravity and string theory \cite{QG1,QG2,QG3,QG4,QG5,QG6}; a similar analysis for generalized models of two matrices is under development. In NRMT, the evolution of the system towards the critical point has been given a clear hydrodynamic interpretation 
\cite{US1,US2,US3,US4,US5,US6,US7} in terms of  the celebrated 
Hele-Shaw free-moving boundary problem 
\cite{HS1,HS2,HS3,HS4,HS5,HS6,HS7,HS8}.  In this paper, we continue this analysis and explore the scaling behavior of the ensemble at a critical point.

\section{Normal Random Matrix Ensembles}

\paragraph{Definitions}

We briefly recall the basic concepts of NRMT \cite{NRM1,NRM2,NRM3,US3}: the 
object of study is the ensemble of $N \times N$ normal matrices $M$
($i.e.$ which commute with their Hermitian conjugates, $[M, M^{\dag}]=0$), 
with statistical weight given by 
\begin{equation}
\label{ZN}
e^{\frac{1}{\hbar}{\rm tr} \, W(M, M^{\dag})} d\mu (M),
\end{equation}
and we choose for the present work the function $W$ to be of the form
\begin{equation}
\label{potential}
W=-|z|^2+V(z)+\overline{V(z)},
\end{equation}
where $V(z)$ is a holomorphic function in a domain which
includes the support of eigenvalues. In (\ref{ZN}), 
$\hbar$ is an area parameter, and the measure of integration
over normal matrices is induced by the flat metric
on the space of all complex matrices. 
Upon integrating out angular degrees of freedom, the joint probability distribution of eigenvalues $z_1,\dots,z_N$ is expressed as
\begin{equation}
\label{mean}
\frac{1}{N! \tau_N}|\Delta_N (z)|^2 \,\prod_{j=1}^N
e^{\frac{1}{\hbar}W(z_j,\bar z_j)} d^2 z_j
\end{equation}
Here
$d^2 z_j \equiv dx_j \, dy_j$ for $z_j =x_j +iy_j$,
$\Delta_N(z)=\det (z_{j}^{i-1})_{1\leq i,j\leq N}=
\prod_{i>j}^{N}(z_i -z_j)$
is the Vandermonde determinant, and
\begin{equation} \label{tau}
\tau_N = \frac{1}{N!}\int
|\Delta_N (z)|^2 \,\prod_{j=1}^{N}e^{\frac{1}{\hbar}
W(z_j,\bar z_j)} d^2z_j
\end{equation}
is a normalization factor, the partition function
of the matrix model.

\begin{figure} \begin{center} 
\includegraphics*[width=5cm]{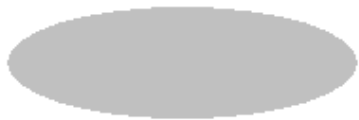}
\caption{The distribution of eigenvalues for the Gaussian potential. The droplet is an ellipse with quadrupole moment $2|t_2|$ and area $\pi\hbar N$.}
\label{ellipse}
\end{center} \end{figure}

\paragraph{Droplets of eigenvalues}
It has been known for a long time that in a proper large $N$ limit 
($\hbar\to 0$, $t_0 = N\hbar$ fixed), the eigenvalues of $M$ densely occupy  
a domain $D$ in the complex plane. The first rigorous result was 
obtained by Ginibre \cite{NRM3} in the form of Circular Law, later
generalized by Girko to Elliptic Law \cite{NRM4}, illustrated in 
Figure~\ref{ellipse}. In this case, the function $V(z)$ introduced in
(\ref{potential}) is a quadratic polynomial, and the corresponding 
matrix model is referred to as Gaussian. Deformations of the droplet
may be introduced by adding higher order terms to the function $V(z)$, 
provided they are small compared to the unperturbed potential, in a 
domain including the droplet \cite{NRM2}. In the region of validity, 
a power expansion of $V(z) = \sum_{k \ge 1} t_kz^k$ is expressed through
the exterior harmonic moments of the droplet of eigenvalues (\ref{moments})
\cite{US5,US6}.  

\paragraph{Wave functions and complex orthogonal polynomials}\la{1A}
In this section we specify the potential (\ref{potential}) such
that it properly defines a scalar product for analytic functions in the sense of Bargmann \cite{B1,B2}. The wave functions and orthogonal polynomials are defined through
\be \la{chin}
\psi_n (z)=
e^{-\frac{|z|^2}{2\hbar}}\chi_n(z),\quad\mbox{and}\quad\chi_n (z)=
e^{\frac{1}{\hbar}V(z)}P_n (z),
\ee
where the holomorphic polynomials $P_n(z)$ are orthogonal in the complex plane with weight $e^{-[|z|^2 - 2\mathcal{R}e V(z)]/\hbar}$. They
obey a set of differential equations with
respect to the argument $z$, and recurrence relations with respect to
the degree $n$. Multiplication by $z$ may be represented in the
basis of $\chi_n$ through a semi-infinite lower triangular matrix 
with one adjacent upper diagonal, $L_{nm}=0$ for $m>n+1$:
\begin{equation}\la{L1}
L_{nm}\chi_m(z)=z\chi_n(z)
\end{equation}
(summation over repeated indices is implied), and the differentiation $\p_z$ is represented by an upper triangular matrix with one adjacent lower diagonal. Integrating by parts the matrix elements of 
$\p_z$, we have: 
\begin{equation}\la{M}
(L^{\dag})_{nm}\chi_m =
\hbar\p_z\chi_n,
\end{equation}
where $L^{\dag}$ is the Hermitian conjugate operator.

Operators $L, \, L^{\dag}$, may also be represented in the basis of the shift operator $\mathcal W$ defined through $\mathcal W
\chi_n=\chi_{n+1},$ and become
\begin{equation}\la{M11}
L=R \mathcal W+\sum_{k\geq 0} U^{(k)} \mathcal W^{-k},\quad
L^{\dag} = \mathcal W^{-1} \bar R +
\sum_{k\geq 0} \mathcal W^{k} \bar U^{(k)},
\end{equation}
where $R, U^{(k)}$ are diagonal semi-infinite matrices of elements $R_{nm} = r_n \delta_{nm},$  $U^{(k)}_{nm} = u^{(k)}_n \delta_{nm}, \, n,m \ge 0,$ and $\delta_{nm}$ is the Kronecker symbol. Acting on $\chi_n$, we obtain the commutation
relation (the ``the string equation", compatibility of Eqs. (\ref{L1}) and (\ref{M}))
\begin{equation}\la{string}
[L^{\dag}, L]=\hbar.
\end{equation}
The string equation provides relations between the coefficients
$r_n$ and $u_n^{(k)}$. In particular, its diagonal part reads
\be\la{Area}
n\hbar=r_n^2-\sum_{k\geq 1}\sum_{p=1}^k|u_{n+p}^{(k)}|^2.
\ee

\paragraph{Example: Normal Gaussian Ensemble}
\la{classicalellipse}

For the case of Gaussian potential $V(z)= t_2 z^2$, $2|t_2|<1$, orthogonal
polynomials $P_n$ are given by complex Hermite polynomials of scaled variable \cite{NRM5}. Recurrence relations
(\ref{M11}) become two-term: 
\be\la{59}
z\psi_n=r_n\psi_{n+1}+u_n\psi_{n-1},\quad\quad
(L^{\dag}\psi)_n=r_{n-1}\psi_{n-1}+\bar u_{n+1}\psi_{n+1},
\ee
and equations (\ref{string}, \ref{Area}) give the parameters $r_n, u_n$ 
recursively through
\be\la{58}
2t_2=\frac{\bar u_{n+1}}{r_n},\quad n\hbar=r_n^2-|u_{n+1}|^2,
\ee
where the second equation is the area formula (\ref{Area}). Defining 
the two-component vector $\underline\chi_n=(\chi_{n-1},\,\chi_{n})^t$, 
we may express the action of the shift operator on $\underline\chi_n$ 
as
$$
{\mathcal W}_{n}=
\left (
\begin{array}{cc}
0 & 1  \\
-\frac{u_n}{r_n}  &
\frac{z}{r_n}
\end{array}
\right).
$$
The operator $L^{\dag}$ also acquires a reduced $2 \times 2$ matrix representation:
\be\la{601}
{\mathcal L}_n \underline\chi_{n+1}=
\left (
\begin{array}{cc}
     z\frac{ r_{n-1}}{u_n} & \bar u_{n+1} - \frac{r_n r_{n-1}}{u_n}  \\
     r_n- \frac{\bar u_{n+2}u_{n+1}}{r_{n+1}} &
\frac{\bar u_{n+2}}{r_{n+1}}z
\end{array}
\right)\underline\chi_{n+1}.
\ee
With the help of (\ref{58}),  ${\mathcal L}_n$ reads
$$
{\mathcal L}_n (z)=
\left (
\begin{array}{cc}
z(2\bar t_2)^{-1} & -r_n(2\bar t_2)^{-1}(1-4|t_2|^2)  \\
r_n(1-4|t_2|^2)  & 2t_2 z
\end{array}
\right).
$$

\paragraph{Complex curve}
Since the operator $L^{\dag}$ represents multiplication by $\bar z$, it 
is possible to define for each value of $n$ a complex curve $f_n(z,\tilde{z}) = 0$ by solving the eigenvalue equation $\det [{\mathcal L}_n (z) - \tilde z] = 0$ and setting $\tilde z = \bar z$; we obtain an ellipse of equation
\be\la{curveellipse}
z\bar z - ( t_2z^2+ \bar t_2\bar z^2)
\frac{2}{1+4|t_2|^2}-n\hbar\frac{1-4|t_2|^2}{1+4|t_2|^2}=0,
\ee
with the quadrupole moment $2|t_2|$ and area $\pi t_0=\pi n\hbar$. 
This is
 precisely the boundary of the domain $\mathbb{D}_n$ filled by eigenvalues of the matrix model, at fixed 
$\hbar, t_2$ and $n$ (or normalized area $t_0$).

\begin{figure}
\begin{center}
\includegraphics*[width=5cm]{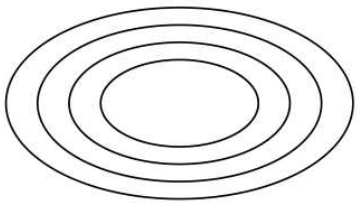}
\caption{Growth in the non-critical Ginibre-Girko ensemble. The
shape parameter of the ellipse  $|t_2| < 1/2$, so the normalized 
area $t_0 = n\hbar$ may be increased indefinitely without encountering any singularity.}
\label{ell}
\end{center} \end{figure}

The geometrical meaning of the complex curve (\ref{curveellipse}) is
 straightforward: at fixed shape parameter $t_2$ and area parameter $\hbar$, 
increasing $n$ yields 
elliptical domains that represent the support of 
the corresponding $n \times n$ Gaussian model. 
A remarkable feature of
this process (represented in Figure~\ref{ell} and labeled $growth$ in our
 previous works \cite{US5,US6,US7}) 
is that it preserves the external harmonic moments of the domain  $\mathbb{D}_n$, 
\beq \la{moments}
t_k(n) = t_k(n-1), \quad t_k(n) = -\frac{1}{\pi k}\int_{\mathbb{C} \setminus \mathbb{D}_n} \frac{d^2 z}{z^k}, \quad k \ge 1.
\eeq
The only harmonic moment which changes in this process is the normalized area  $t_0 = \frac{1}{\pi} \int d^2 z,$
and it increases in increments of $\hbar$ (hence the meaning of $\hbar$ as quantum of area).  We may say that the growth 
of the NRM ensemble consists of increasing the area of the domain by multiples of $\hbar$, while preserving all the other
external harmonic moments.  The continuum version of this process, known as $Laplacian$ $Growth$, is a famous 
problem of complex analysis. It arises in the two-dimensional hydrodynamics of  two non-mixing fluids, 
one inviscid and the other viscous, upon neglecting the effects of surface tension (the Hele-Shaw problem 
\cite{HS1,HS2,HS3,HS4,HS5,HS6,HS7,HS8}). 

\paragraph{Deformations and critical points of the ensemble}

\begin{figure}
\begin{center}
\includegraphics*[width=5cm]{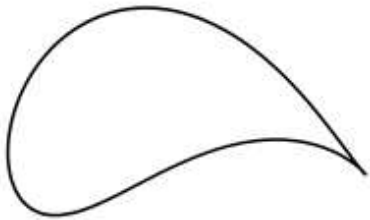}
\caption{Deformation of the Gaussian ensemble and formation of boundary singularities.}
\label{jouk}
\end{center} \end{figure}

One of the most important properties of Laplacian Growth is that,  
with the exception of special choices for the external harmonic moments 
$t_k$, the growth ends at a finite (critical) value of the area $t_0 = t_c$,
when cusp-like singularities form on the boundary of the domain. 

Laplacian Growth of a simply-connected domain $\mathbb{D}(t_0)$ can be described \cite{US1,US2,US3} through a conformal map $z(w, t_0)$ which takes $\mathbb{C} \setminus \mathbb{B}_1$ into $\mathbb{C} \setminus \mathbb{D}(t_0)$, where $\mathbb{B}_1$ is the unit disk centered at the origin in the $w-$plane. Therefore, the conformal map has the form
\be
z(w, t_0) = r(t_0)w + \sum_{k \ge 0}u_k(t_0)w^{-k},
\ee
and the coefficients are functions of time (normalized area) $t_0$, such 
that 
\be \la{poisson}
\{z(w, t_0), \bar z(w^{-1}, t_0) \} \stackrel{def}{=} 
w \left [ \frac{\p z}{\p w}\frac{\p \bar z}{\p t_0} - 
\frac{\p z}{\p t_0}\frac{\p \bar z}{\p w} \right ] = 1.
\ee
The Poisson bracket (\ref{poisson}) encodes the infinite number of conservation laws
\be
\frac{d t_k}{d t_0} = 0, \quad t_k(t_0) = -\frac{1}{\pi k}\int_{\mathbb{C} \setminus \mathbb{D}(t_0)} \frac{d^2 z}{z^k},
\ee
as well as the classical area formula
$$
t_0 = r^2 -\sum_{k}k |u_k|^2.
$$
As mentioned in the previous paragraph, we may regard Laplacian Growth 
as the continuum limit of a corresponding NRM ensemble, sharing the same
set of exterior harmonic moments. For instance, choosing the function 
$V(z)$ of the form 
\be \la{logg}
V(z) = \alpha \log \left ( \frac{\beta}{\beta - z} \right ) =  \sum_{k \ge 1}
\frac{\alpha }{k\beta^k} z^k, \quad \mathcal{R}e (\alpha) < 0,
\ee
corresponds to harmonic moments $t_k  = \frac{\alpha}{k \beta^k}$ 
and the exterior of the droplet is given by the map \cite{US5}
\be \la{jmap}
z(w) = w\left [r + \frac{u}{a(w-a)} \right ], \quad |w| \ge 1, 
\ee
where
\be
\beta =\frac{r}{\bar a} + \frac{u}{a(1-|a|^2)}, \quad
t_0 = \bar \alpha + r^2 - \frac{ur}{a^2}.
\ee
For suitable values of $\alpha$, the orthogonal polynomials corresponding
to (\ref{logg}) are well defined, though they may obey complicated recurrence relations. In the continuum limit of the model, the droplet
grows until its area reaches the critical value and a cusp forms on the
boundary, Figure~\ref{jouk}. Formation of critical points is best 
described using the complex curve associated with the conformal map 
(\ref{jmap}). As indicated in \cite{US5}, it is a degenerate elliptic 
curve, with two branch points at $z_{1,2}$ inside the domain and a double point $z_*$ outside. The critical point on the boundary appears when one of the branch points  inside merges with the double point, leading to a cusp expressed in local coordinates as $x^3 \sim y^2$.  Restoring finite values for $n$ and $\hbar$ is equivalent to a discretization of the Laplacian Growth and lifts the degeneracy of the complex curve of the continuum limit \cite{US6,US7}.

\paragraph{Universality in the scaling region at critical points: a conjecture}

\begin{figure}
\begin{center}
\includegraphics*[width=6cm]{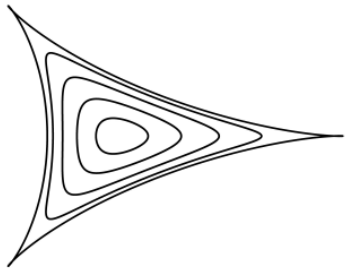}
\caption{Critical points obtained from the cubic potential 
$V(z)=t_3z^3$. Area of the droplet is increased at fixed
$t_3$ until it reaches a maximal, critical value.}
\label{hyp}
\end{center} \end{figure}

Detailed analysis of critical Hermitian ensembles indicates that the 
behavior of orthogonal polynomials in a specific region including the
critical point (the {\it {scaling region}}), upon appropriate scaling
of the degree $n$, is essentially independent of the bulk features of the ensemble. This $universality$ property (a common working hypothesis in the physics of critical phenomena) is expected 
to occur for critical NRM ensembles as well -- and is indeed easy to 
verify in critical Gaussian models, $2|t_2|=1$. Analytically, it means that by suitable
scaling of the variables $z, n$: 
$$
n \to \infty, \,\,\, \hbar \to 0, \,\,\, n\hbar = t_0, \,\,\,
t_0 = t_c - \hbar^{\delta} \nu, \,\,\, z = z_c + \hbar^{\epsilon} \zeta,
$$  
where $z_c$ is the location of the critical point and $t_c$ the critical
area, the wave function $\Psi_n(z)$ will reveal a universal part 
$\phi(\nu,\zeta)$ which depends exclusively on the local singular 
geometry $x^p \sim y^q$ ($p, q$ mutual primes) of the complex curve at the critical point. This conjecture is a subject of active research. Its main consequence is that in order to describe the 
scaling behavior for a certain choice of $p, q$, 
it is possible to replace a given ensemble with another which leads to
the same type of critical point, though they may be very different at 
other length scales. More precisely, the scaling behavior of operators in the vicinity of singular points illustrated in Figure \ref{jouk} is assumed to be identical to that of singular points in Figure \ref{hyp} (where 
$V(z) =t_3z^3$), although the two critical droplets are obtained starting from different potentials. A constructive argument for this method is under 
development \cite{myself}. 

\section{Scaling at critical points of normal matrix ensembles} \label{painl}

\begin{figure}
\begin{center}
\includegraphics*[width=7cm]{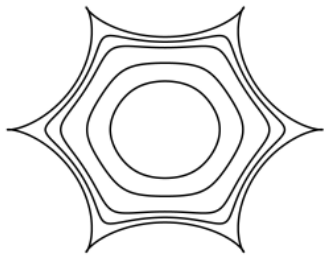}
\caption{Critical boundary corresponding to $V(z)=t_6z^6$.}
\label{six}
\end{center} \end{figure}

In the remainder of the paper we analyze the regularization of Laplacian
Growth for a critical point of type $p=3, q=2$, by discretization of the 
conformal map as described in the previous section. For simplicity, we
start from the conformal map corresponding to the potential $V(z) = t_3z^3$, which is the simplest model leading to the specified type of cusp. It 
should be noted that the analysis will be identical for any monomial potential $V(z) = t_nz^n, n \ge 3$; for every such map, $n$ singular 
points of type $p=3, q=2$ will form simultaneously on the boundary. The 
critical boundary corresponding to $n=6$ is shown in Figure~\ref{six}.

\paragraph{Painlev\'e I as string equation}

We start from the Lax pair corresponding to the potential $V(z)=t_3z^3$ 
\cite{US5}:
\be
L\psi_n = r_n \psi_{n+1} + u_n \psi_{n-2}, \,\, 
L^{\dag}\psi_n = r_{n-1}\psi_{n-1} + \bar u_{n+2}\psi_{n+2}.
\ee
The string equation (\ref{string}) $[L^{\dag}, \, L] = \hbar$ 
translates into 
\begin{eqnarray} \nonumber
(r^2_n +|u|_n|^2 - r^2_{n-1} - |u|_{n+2}^2) \psi_n  +  (r_n\bar u_{n+3} - r_{n+2}\bar u_{n+2}) \psi_{n+3} \\
+(r_{n-3}u_n - r_{n-1}u_{n-1}) \psi_{n-3} = \hbar \psi_n.
 \end{eqnarray}
Identifying the coefficients gives
\be \la{ar}
\left ( r_n^2 - |u|^2_{n+2} - |u|^2_{n+1} \right ) - \left ( r_{n-1}^2 - |u|^2_{n+1} - |u|^2_{n} \right ) = \hbar, 
\ee
and
\be \la{co}
\frac{\bar u_{n+2}}{r_nr_{n+1}} = \frac{\bar u_{n+3}}{r_{n+1}r_{n+2}} = 3t_3.
\ee
Equation (\ref{ar}) gives the quantum area formula
\be
r_n^2 - (|u|_{n+2}^2 + |u|_{n+1}^2) = n\hbar,
\ee
which together with the conservation law (\ref{co}) leads to the discrete Painlev\'e equation
\be
r_n^2 \left [ 1 - 9|t_3|^2 (r_{n-1}^2  + r_{n+1}^2) \right ] = n\hbar.
\ee

\noindent In the continuum limit, the equation becomes
\be
r^2 - 18|t_3|^2 r^4 = t_0.
\ee
The critical (maximal) area is given by 
\be
\frac{dt_0}{dr^2} = 0, \,\,  36|t_3|^2r^2_c = 1. 
\ee
Choosing $r_c =1$ gives $6|t_3| = 1$ and $t_c = \frac{1}{2}$. It also follows that
\be
u_n = \frac{r_{n-2}r_{n-1}}{2}, \,\, z_c = \frac{3}{2}.
\ee
Introduce the notations
\be
N\hbar = t_c, \,\, n\hbar = t_0 = t_c +\hbar^{4a} \nu, \,\, r_n^2 = 1 - \hbar^{2a}u(\nu) , 
\,\, z = \frac{3}{2} + \hbar^{2a}\zeta,    
\ee
where $a=\frac{1}{5}.$ We get $\p_n = \hbar^a \p_\nu$ and
\be
r^2_{n+k} = 1 - \hbar^{2a}u - k \hbar^{3a}\dot u(\nu) - \frac{k^2}{2}\hbar^{4a}\ddot{u}, 
\ee
where dot signifies derivative with respect to $\nu$. 
The scaling limit of the quantum area formula becomes
\be
(1-\hbar^{2a}u)\left [\frac{1}{2} + \hbar^{2a}\frac{u}{2} + \hbar^{4a} \frac{\ddot u}{4} \right ] = \frac{1}{2} + \hbar^{4a}\nu,
\ee
giving at order $\hbar^{4a}$ the Painlev\'e equation
\be \la{painleve}
\ddot u - 2 u^2 = 4 \nu.
\ee
Rescaling $u \to c_2 u$, $\nu \to c_1 \nu$ gives the standard form
\be
\ddot u - 3u^2 = \nu,
\ee
for $c_2 = 4c_1^3, 8c_1^5 = 3$.

\paragraph{Painlev\'e as compatibility equation}

Use the modified wave functions ($\mbox{Pol}$ 
represents the polynomial part)
\be
\phi_n = \prod_{i=0}^{n-1}r_i \psi_n, \,\, 
\mbox{Pol } \phi_n(z) = z^n + O(z^{n-1}),
\ee
and rewrite the equations for the Lax pair as
\be
L \phi_n = \phi_{n+1} + \frac{r^2_{n-2}r^2_{n-1}}{2} \phi_{n-2}, \,\, 
L^{\dag} \phi_n = r^2_{n-1}\phi_{n-1} + \frac{\phi_{n+2}}{2}.
\ee
Notice that using the shift operator $\mathcal{W}$, the system can also be written
\be
L = \mathcal{W} + \frac{1}{2}\left ( r^2_{n-1}\mathcal{W}^{-1} \right )^2, \,\, 
L^{\dag} = r^2_{n-1}\mathcal{W}^{-1} + \frac{1}{2} \mathcal{W}^2. 
\ee
Introduce the scaling $\psi$  function through 
\be
\phi_n(z) = e^{\frac{z^2}{2\hbar}} \psi(\zeta, \nu).
\ee
The action of Lax operators on $\psi$ gives the representation
\be
L = \frac{3}{2} + \hbar^{2a}\zeta, \, \, L^{\dag} = z + \hbar \p_\zeta = \frac{3}{2} + \hbar^{2a} \zeta + \hbar^{3a} \p_\zeta.
\ee
Therefore, the action of $\zeta$ is given by the sum of equations at order $\hbar^{2a}$:
\be
3 + 2 \hbar^{2a} \zeta  = \mathcal{W} + \frac{1}{2} \mathcal{W}^2 + r^2_{n-1}\mathcal{W}^{-1}
+ \frac{1}{2}\left ( r^2_{n-1}\mathcal{W}^{-1} \right )^2, 
\ee
and the action of $\p_\zeta$ by their difference:
\be
\hbar^{3a}\p_\zeta  = - \mathcal{W} + \frac{1}{2} \mathcal{W}^2 + r^2_{n-1}\mathcal{W}^{-1}
- \frac{1}{2}\left ( r^2_{n-1}\mathcal{W}^{-1} \right )^2. 
\ee
Equivalently, we can write
\be
\hbar^{2a}\zeta = \frac{1}{2} \left [ \left ( \mathcal{W}+1 \right )^2 + 
\left ( r^2_{n-1}\mathcal{W}^{-1} + 1 \right )^2  \right ] - 4,
\ee
\be
\hbar^{3a}\p_\zeta
= \frac{1}{2} \left [ \left ( \mathcal{W}-1 \right )^2 - 
\left ( r^2_{n-1}\mathcal{W}^{-1} - 1 \right )^2  \right ].
\ee
Expanding the shift operator in $\hbar$ leads to
\be
\mathcal{W} = 1 + \hbar^{a} \p_\nu + \hbar^{2a}\frac{\p^2_\nu}{2} + \hbar^{3a}\frac{\p^3_\nu}{6},
\ee
and
\be
r^2_{n-1}\mathcal{W}^{-1} = 1 - \hbar^{a}\p_\nu + \hbar^{2a}\left (\frac{\p^2_\nu}{2}-u \right) + 
\hbar^{3a}\left ( -\frac{\p^3_\nu}{6}  + u \p_\nu + \dot u \right ).
\ee
Substituting into the equations for $\zeta, \p_\zeta$ gives the system of equations
\be
\ddot \psi = \frac{2(\zeta + u)}{3}\psi, \,\,\,\, \psi' = \frac{\dot u}{6} \psi +  \frac{2\zeta - u}{3}\dot \psi, 
\ee
where primed variables are differentiated with respect to $\zeta$. The equations can be written in matrix form as
\be
\Psi ' = \Lambda \Psi, \,\, \dot \Psi = Q \Psi, \,\, \Psi = 
\left (
\begin{array}{c}
\psi \\
\dot \psi
\end{array}
\right ),
\ee
where
\be
\Lambda = 
\left (
\begin{array}{cc}
\frac{\dot u}{6} & \frac{2\zeta - u}{3} \\
\frac{\ddot u}{6} + \frac{2(\zeta + u)(2\zeta - u)}{9} \,\,\,\, & -\frac{\dot u}{6}
\end{array}
\right ), \,\,
Q = 
\left (
\begin{array}{cc}
0 & 1 \\
\frac{2(\zeta + u)}{3} \,\,\, & 0
\end{array}
\right ).
\ee
The compatibility equations 
\be
\dot \Lambda - Q' = [Q, \,\Lambda]
\ee
yield the Painlev\'e equation derived in the previous section:
\be
\dot \Lambda = 
\left (
\begin{array}{cc}
\frac{\ddot u}{6} & -\frac{ \dot u}{3} \\
\frac{\stackrel{\ldots}{u}}{6} + \frac{2\zeta \dot u -4u \dot u }{9} \,\,\,\, & -\frac{\ddot u}{6}
\end{array}
\right ), \,\,
Q' = 
\left (
\begin{array}{cc}
0 & 0\\
\frac{2}{3} \, & 0
\end{array}
\right ),
\ee
and
\be
[Q, \,\, \Lambda ] =
\left (
\begin{array}{cc}
\frac{\ddot u}{6} & -\frac{ \dot u}{3} \\
\frac{2(\zeta+u) \dot u }{9} \,\,\,\, & -\frac{\ddot u}{6}
\end{array}
\right ).
\ee
Thus,
\be
0 = \dot \Lambda - Q' - [Q, \,\Lambda] =
\left (
\begin{array}{cc}
0 & 0\\
\frac{\stackrel{\ldots}{u}}{6} - \frac{6u \dot u}{9} - \frac{2}{3} \,\,\,\,\,\, & 0
\end{array}
\right ).
\ee
The only non-trivial element of the matrix gives
\be
\stackrel{\ldots}{u} -4u\dot u - 4 = 0,
\ee
$i.e.$ the Painlev\'e equation derived in the previous section.

\paragraph{Painlev\'e and the non-degenerate spectral curve}

\begin{figure}
\begin{center}
\includegraphics*[width=6cm]{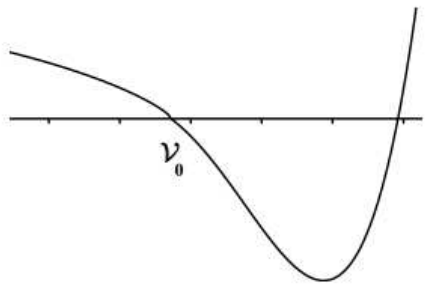}
\caption{Physical solution for Painlev\'e I, with no singularities 
along the negative real axis.}
\label{pain}
\end{center} \end{figure}

\begin{figure}
\begin{center}
\includegraphics*[width=12cm]{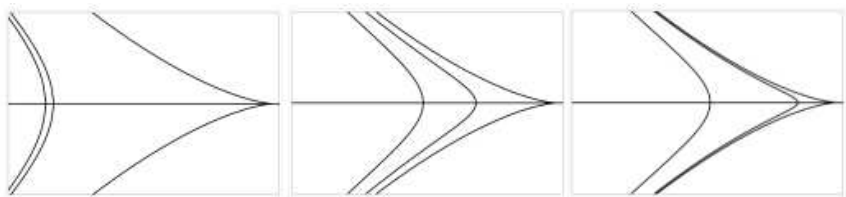} \end{center}
\caption{Evolution of the non-degenerate curve (the leftmost boundary in each diagram) and degenerate curve (middle boundary) relative to the singular curve (right boundary), at (a) $\nu \to -\infty$, (b) $\nu_0 - \nu = O(1)$ and (c) $\nu_0 - \nu \ll \nu_0$.}
\label{last}
 \end{figure}

In the scaling limit, the spectral curve is defined by the eigenvalue equation for the operator $\p_\zeta$, 
\be
\p_\zeta \Psi = \lambda(\zeta) \Psi,
\ee
or
\be
\det 
[\lambda(\zeta) - \Lambda] =0.
\ee
We can write it also explicitly as an elliptic curve,
\be 
\lambda^2 = \left ( \frac{2\zeta}{3} - \frac{u}{3}
\right )^2 \left (\frac{2\zeta}{3} + \frac{2u}{3} \right )
+
\frac{2\zeta}{3}\cdot \frac{\ddot u}{6} + 
\frac{\dot{u}^2 - 2u \ddot u}{36}.
\ee
The critical points of the curve solve
\be
\lambda_{+} = \lambda_{-} = 0,
\ee
or 
\be \la{curve}
\mu^3 +\frac{2\nu}{3} \mu + \left [ 
\left (\frac{\dot u}{6} \right )^2 - \frac{(u^2+6\nu)u}{27} \right ] = 0,
\ee
where $\mu = 2\zeta/3$. Setting all derivatives to zero in (\ref{painleve})  and  ($\ref{curve}$), we get the degenerate solutions
\be
\zeta_{1} = -u = -\sqrt{-2\nu}, \,\, \zeta_{2,3} = \frac{u}{2} = \frac{\sqrt{-2\nu}}{2}.
\ee
We choose (up to exponential corrections) the solution to Painlev\'e
I which is free of poles along the negative real axis, Figure \ref{pain}, and follow the evolution of the boundary as $t_0 \to t_c$, 
Figure \ref{last}. As one can see, the presence of derivative terms
lifts the degeneracy of the curve, so that the boundary remains
smooth even as the area reaches its critical value.

\subparagraph{(1) Quantum curve at $\nu \to -\infty$}

In the case $\nu \to -\infty$, the asymptotic expansion of the solution to Painlev\'e equation reads
\be
u = \sqrt{-2\nu} - \frac{16}{\nu^2} + O(\nu^{-3}),
\ee
subject to exponential corrections. The curve becomes 
(Figure \ref{last}, first diagram) nondegenerate, with simple critical points
\be
\zeta_1 = - \sqrt{-2\nu} -\frac{1}{96\nu^2}, \,\,
\zeta_{2,3} = \frac{\sqrt{-2\nu}}{2} + \frac{1}{192\nu^2} \pm \frac{i}{(-2\nu)^{3/4}}.
\ee

\subparagraph{(2) Quantum curve at $u = 0$}

Let $u(\nu_0) = 0$ define the smallest real solution of $u(\nu)=0$. Then the local expansion reads
\be
u(\nu) = \alpha (\nu -\nu_0) + 2\nu_0 (\nu -\nu_0)^2
+\frac{2}{3}(\nu -\nu_0)^3 + \frac{\alpha^2}{6}(\nu -\nu_0)^4 + \dots
\ee
Merging this regular expansion with the asymptote at $\nu \to -\infty$ yields
\be
\nu_0 = -2^{\frac{7}{5}}, \,\, \alpha = \frac{1}{8\nu_0^3} + \frac{1}{\sqrt{-2\nu_0}} = -3\cdot2^{\frac{2}{5}}.
\ee
The discriminant of (\ref{curve}) at $\nu=\nu_0$ becomes
$
\Delta^2 = \frac{9\alpha^4 + (8\nu_0)^3}{6^6} > 0,
$
so the equation has again one real solution and two complex conjugate roots. Moreover, since the free term is positive, it follows that the real solution is negative. The physical interpretation shows that the curve is smoothed out (Figure \ref{last}, diagrams b and c) at $u=0$, instead of forming the classical $(2,3)$ cusp given by the degenerate curve.

\section*{Acknowledgments}

The author is indebted to P. Wiegmann and A. Zabrodin for help, suggestions and advice. Very beneficial discussions  with I. Krichever are gratefully acknowledged. Special thanks are owed to I. Aleiner and A. Millis at Columbia University for support, and J. Harnard and M. Bertola for the  stimulating research environment at the Centre for Mathematical Research, University of Montreal, where this work was presented. Relevant suggestions 
from reviewers were very helpful in clarifying certain aspects of the formalism used in this work.


\section*{References}

\begin{thebibliography}{10}

\bibitem{WD1} 
Wigner E~P 1951 {\it {Ann. of Math.}} (2)  {\bf{53}} 36-67 

\bibitem{WD2} 
Dyson F 1962 {\it {J. Math. Phys.}} {\bf {3}} 140-156

\bibitem{2HRM1}
Bertola M, Eynard B  and  Harnad J 2003 {\it{Theor. Math. Phys.}} {\bf {134}} 27-38 

\bibitem{2HRM2}
Bleher P and Its A 2002 {\it {math-ph/0201003}} 

\bibitem{2HRM3}
Bertola M, Eynard B  and  Harnad J 2003 {\it {J. Phys. A}} 
{\bf {36}} 3067-3084

\bibitem{2HRM4}
Bertola M, Eynard B  and Harnad J 2003 {\it {Comm. Math. Phys.}} 
{\bf {243}} no.2 193-240

\bibitem{2HRM5}
Kapaev A 2003 {\it {J. Phys. A}} {\bf 36} 4629-4640

\bibitem{2HRM6}
Bleher P and Its A. 2004 {\it {math-ph/0409082}}

\bibitem{2HRM7}
Bertola M, Eynard B  and Harnad J 2004 {\it {nlin.SI/0410043}}

\bibitem{US1} Mineev-Weinstein M, Wiegmann P~B and
Zabrodin A 2000 {\it {Phys. Rev. Lett.}} {\bf {84}} 5106

\bibitem{US2}
Kostov I~K,  Krichever I,   Mineev-Weinstein M,
Wiegmann P~B and Zabrodin A 2001  MSRI {\bf{ 40}}
285 (Cambridge: Cambridge Univ. Press.) 

\bibitem{US3}
Wiegmann P~B and Zabrodin A	2003 {\it {J. Phys. A}} 
{\bf {36}} 3411-3424

\bibitem{US4}
Agam O, Bettelheim E, Wiegmann P~B
and Zabrodin A 2002 {\it {Phys. Rev. Lett.}} 
{\bf{88}} 236802

\bibitem{US5}
Teodorescu R, Bettelheim E,  Agam O, 
Zabrodin A  and Wiegmann P 2005 
{\it {Nucl. Phys.}}  {\bf {B704}} 407; ibid  2004 {\bf{700}}  521 

\bibitem{US6}
Teodorescu R, Wiegmann P and Zabrodin A 2005 
{\it {Phys. Rev. Lett.}} {\bf{95}} 044502 

\bibitem{US7}
Bettelheim E, Wiegmann P
and Zabrodin A 2005 {\it {arXiv.org:nlin/0505027}} 


\bibitem{CM1}Wiegmann P~B and Zabrodin A 2000
{\it {Comm. Math. Phys.}} {\bf {213}}  523-538

\bibitem{CM3}
Marshakov A, Wiegmann P~B and Zabrodin A 2002 {\it {Comm. Math. Phys.}} {\bf {227}}  1 131-153

\bibitem{CM4}
Krichever I, Marshakov A and Zabrodin A 2003  {\it {hep-th/0309010}}
	
\bibitem{QG1}
Fokas A~S, Its A~R and Kitaev A 1992 {\it {Comm. Math. Phys.}}
 {\bf {147}} 395-430

\bibitem{QG2}
David F 1993 {\it {Phys. Lett.}} {\bf {B302}} 403-410; {\it {hep-th/9212106}}

\bibitem{QG3}
Di Francesco P, Ginsparg P and Zinn-Justin J 1995
{\it {Phys. Rept.}} {\bf {254}} 1-133

\bibitem{QG4}
Chekhov L  and   Mironov A 2002 {\it {hep-th/0209085}}

\bibitem{QG5}
Kazakov V~A and Marshakov A 2003 {\it {J. Phys. A}} {\bf {36}} 3107-3136

\bibitem{QG6}
Dijgraaf R and Vafa C 2002 {\it {hep-th/0208048, hep-th/0206255, hep-th/0207106, hep-th/0302011}}

\bibitem{HS1}
Hele-Shaw H~S~S 1898 {\it {Nature}}

\bibitem{HS2}
Galin L~A 1945 {\it {Dokl. Akad. Nauk SSSR}} {\bf {47} }  250-253

\bibitem{HS3}
Polubarinova-Kochina P~Ya 1945 {\it {Dokl. Akad. Nauk SSSR}} {\bf {47}} 254-257

\bibitem{HS4}
Kufarev P~P 1947 {\it {Dokl. Akad. Nauk SSSR}}
{\bf {57} }  335-348

\bibitem{HS5}
Howison S,  Lacey A  and  Ockendon J 1985  
{\it {Q. J. Mech. Appl. Math.}} {\bf {38}} 343

\bibitem{HS6}
Howison S 1986 {\it {SIAM J. Appl. Math.}} {\bf {46}} 20

\bibitem{HS7}
Hohlov Y and  Howison S 1993 {\it {Quart. Appl. Math.}} {\bf {51}} 777

\bibitem{HS8}Bensimon D,  Kadanoff L~P,
Liang S, Shraiman B~I and Tang C 1986
{\it {Rev. Mod. Phys.}} {\bf {58}} 977


\bibitem{NRM1}Chau L and Zaboronsky O 1998 {\it {Comm. Math. Phys.}} 
{\bf {196}} 203--247 

\bibitem{NRM2}Elbau P and Felder G 2004 {\it {math/0406604}}

\bibitem{NRM3}Ginibre J 1965 {\it {J. of Math. Phys.}} 
{\bf {6}} (3)  440

\bibitem{NRM4}Girko V~L 1986 {\it {Theory of Probability and
Its Applications}} {\bf {30}} (4) 677-690

\bibitem{B1}
Bargmann V 1961 {\it{Comm. Pure Appl. Math.}} {\bf{14}} 187--214

\bibitem{B2}
Bargmann V 1962 {\it{Proc. Nat. Acad. Sci. U.S.A.}} {\bf{48}} 199--204

\bibitem{Orlov}Orlov A Yu 2005 {\it {Acta Appl. Math.}} {\bf {86}}
131-158

\bibitem{NRM5} Di Francesco P, Gaudin M, Itzykson C and Lesage P 1994
{\it {hep-th/9401163}}; Akemann G 2002 {\it {J. Phys. A}} {\bf {36}} 3363

\bibitem{myself} Teodorescu R {\it {unpublished}}.

\end{thebibliography}
\end{document}